\documentclass[iop]{emulateapj}
\usepackage{hyperref}
\usepackage{natbib}
\usepackage{amsmath,amstext}
\usepackage[T1]{fontenc}
\usepackage[figure,figure*]{hypcap}
\usepackage{threeparttable}

\begin{document}

\title{Statistical Analysis of Hubble/WFC3 Transit Spectroscopy of Extrasolar Planets}

\author{Guangwei Fu}
\author{Drake Deming}
\affiliation{Department of Astronomy, University of Maryland, College Park, MD 20742, USA; gfu@astro.umd.edu}
\author{Heather Knutson}
\affiliation{Division of Geological and Planetary Sciences, California Institute of Technology, Pasadena, CA 91125, USA}
\author{Nikku Madhusudhan}
\affiliation{Institute of Astronomy, University of Cambridge, Madingley Road, Cambridge CB3 0HA, UK}
\author{Avi Mandell}
\affiliation{Planetary Systems Laboratory, NASA's Goddard Space Flight Center, Greenbelt MD, 20771, USA}
\author{Jonathan Fraine}
\affiliation{Space Telescope Science Institute, 3700 San Martin Drive, Baltimore MD, 21218, USA}
\begin{abstract}
Transmission spectroscopy provides a window to study exoplanetary atmospheres, but that window is fogged by clouds and hazes.  Clouds and haze introduce a degeneracy between the strength of gaseous absorption features and planetary physical parameters such as abundances.  One way to break that degeneracy is via statistical studies.  We collect all published HST/WFC3 transit spectra for 1.1-1.65\,$\mu$m water vapor absorption, and perform a statistical study on potential correlations between the water absorption feature and planetary parameters. We fit the observed spectra with a template calculated for each planet using the Exo-Transmit code.  We express the magnitude of the water absorption in scale heights, thereby removing the known dependence on temperature, surface gravity, and mean molecular weight.  We find that the absorption in scale heights has a positive baseline correlation with planetary equilibrium temperature; our hypothesis is that decreasing cloud condensation with increasing temperature is responsible for this baseline slope. However, the observed sample is also intrinsically degenerate in the sense that equilibrium temperature correlates with planetary mass. We compile the distribution of absorption in scale heights, and we find that this distribution is closer to log-normal than Gaussian. However, we also find that the distribution of equilibrium temperatures for the observed planets is similarly log-normal.  This indicates that the absorption values are affected by observational bias, whereby observers have not yet targeted a sufficient sample of the hottest planets. 
\end{abstract}
\nopagebreak
\section{Introduction}
Robust observations of exoplanetary atmospheres using transmission and emission spectroscopy with the Wide Field Camera-3 (WFC3) on the Hubble Space Telescope (HST) have led to significant progress in understanding exoplanetary atmospheres (see reviews by \citealp{crossfield15} and \citealp{deming17}).  Recent intriguing results have inferred atmospheric thermal structure and circulation patterns \citep{stevenson14}, temperature inversions \citep{evans17, haynes15}, clouds/hazes \citep{sing16}, and water abundance (\citealp{wakeford17}, \citealp{kreidberg14AJ}). Focusing on HST/WFC3 transmission spectrum measurements, the amplitude of water vapor absorption (1.1 to 1.7\,$\mu$m) has been the most commonly used observational quantity due to its relatively high abundance and strong absorption strength.  One key scientific motivation is to derive the abundance of oxygen (as a proxy for planetary metallicity) as a function of planetary mass \citep{kreidberg14AJ}. The planetary mass-metallicity relation could yield insights into the planet formation process \citep{thorngren16}.
 
However, accurately measuring water abundance through transmission spectroscopy has been argued to be very challenging if only WFC3 spectra are considered (\citealp{griffith14}, \citealp{heng17}). For example, the presence of patchy clouds/hazes can mimic the same effect as either high molecular weight \citep{line16b} or low molecular abundances \citep{madhu14} and also introduce a degeneracy between reference pressure and water abundance in the planetary atmosphere. To optimize and prepare for future transmission spectroscopy observations using the James Webb Space Telescope (JWST), it is important to better understand the effects of clouds and hazes, and develop techniques to precisely measure water abundance in exoplanetary atmospheres. One approach is to perform very in-depth studies of individual planets.  Utilizing additional observational constraints from optical to infrared (0.5 - 5\,$\mu$m), combined with detailed modeling of T-P profiles, properties of cloud-forming condensate species can be deduced (\citealp{wakeford17}, \citealp{line16a}, \citealp{stevenson17}, \citealp{macdonald17}). Once a large sample of planets have been analyzed extensively, patterns and correlations between water abundance and cloud properties may emerge \citep{sing16}.
 
Another approach is to short-circuit the tedious process of multiple in-depth investigations, by seeking correlations between the observed magnitude of water absorption and bulk properties of the exoplanets such as equilibrium temperature, planetary mass and gravity. This approach can also help to formulate hypotheses and reveal potentially related physical parameters that can be tested by subsequent analyses and observations.
 
Recently, \citet{tsiaras17} announced a catalog of hot-Jupiter absorption spectra observed in multiple programs using HST/WFC3 in spatial scanning mode. In this Letter we use these spectra with 4 additional spectra (\citealp{huitson13}, \citealp{kreidberg14N}, \citealp{knutson14b}, \citealp{mandell13}) in a statistical analysis of transit water absorption in relation to planetary bulk parameters for a sample of 34 hot-Jupiter (to hot-Neptune) exoplanets.  Our analysis uses public data and models, and simple techniques that anyone can reproduce. For reasons that we explain below, our fundamental observational quantity is the number of atmospheric scale heights that are opaque in the water band during transit \citep{stevenson16}.  Sec.~2 describes how we determine that quantity based on the spectra from \citet{tsiaras17}, and Sec.~3 describes the correlation of the inferred absorption with other planetary properties.  Sec.~4 summarizes our conclusions and discusses implications for future measurements.   

\begin{figure}
  \includegraphics[width=0.5\textwidth,keepaspectratio]{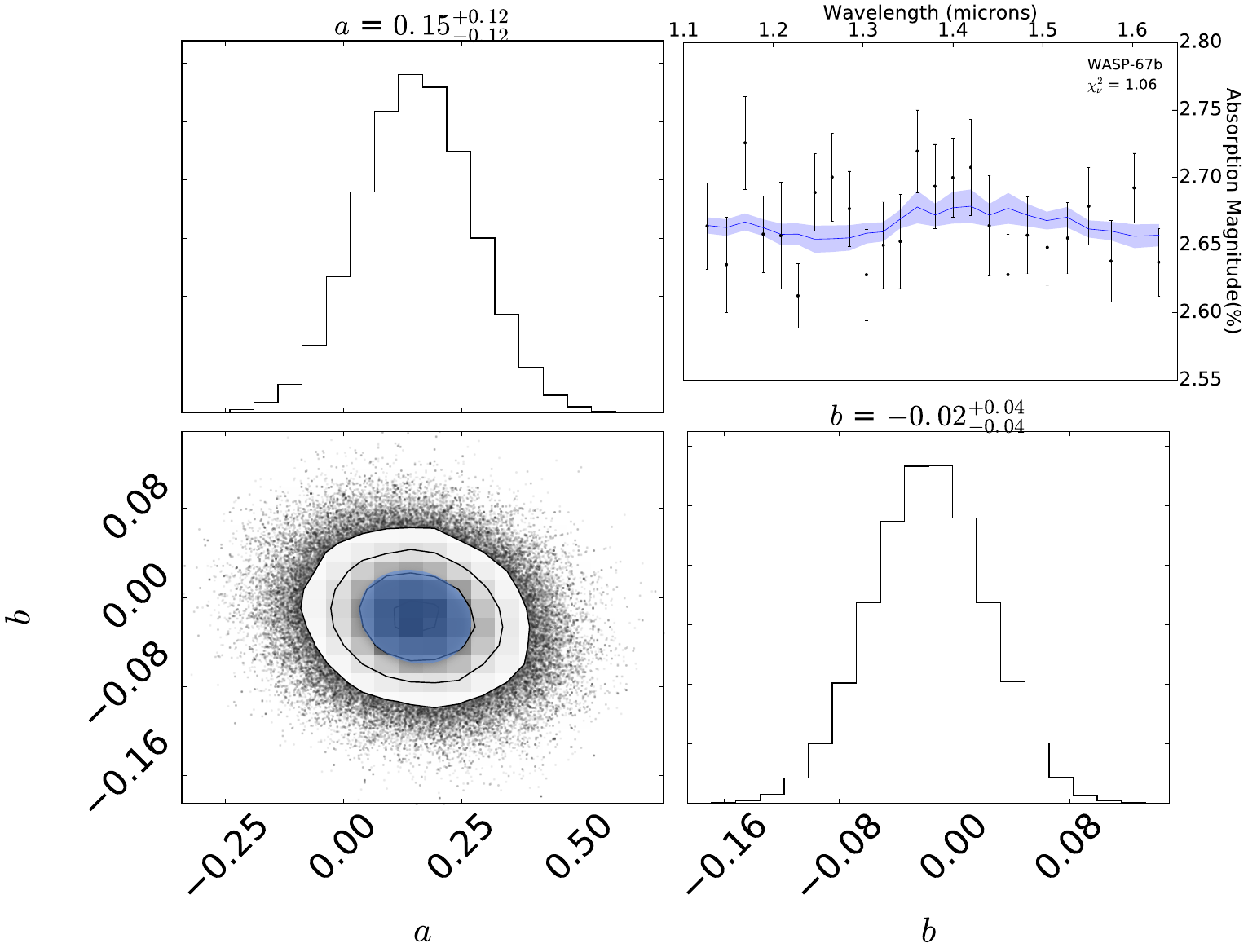}
  \caption{Example of our MCMC fitting of an Exo-transmit template spectrum to the data for WASP-67b. This planet has the median ${\chi_\nu}^2$ in the sample, with ${\chi_\nu}^2 =1.05$.  The $a$ coefficient is the amplitude scale factor for fitting the Exo-transmit template spectrum, and $b$ is the wavelength coefficient of a baseline slope.  The $\le 1\sigma$ contours are shaded in blue for the posterior distribution samples (lower left) and for the fit to the data (upper right). }
  \label{fig:two_panel_fullfit}
\end{figure}

\section{Data Analysis}

A planet with base radius $R_p$ has a transit depth of $R_p^2/R_s^2$ ($R_s$ is star's radius). If the planet's atmosphere is opaque over one scale height, the transit depth will increase by $2R_pH/R_s^2$, where the pressure scale height is $H = kT/\mu g$ with k being the Boltzmann constant, $T$ being the planet's equilibrium temperature, $\mu$ being the mean molecular weight and $g$ being the surface gravity.  Surface gravity and temperature can be directly estimated from measurable quantities including planetary mass and radius, orbital semi-major axis, and stellar temperature and radius (with assumptions on the planetary albedo and longitudinal circulation).  We want to determine how the magnitude of atmospheric absorption varies with physical quantities that are not directly associated with the pressure scale height, such as the existence and height of clouds and hazes. Therefore, following \citet{stevenson16}, we remove the dependence on known parameters by dividing the magnitude of atmospheric absorption by $2R_pH/R_s^2$.  To calculate $H$ and enable a consistent comparison with \cite{stevenson16}, we used a mean molecular weight of 3.8 for planets with $R <$ $0.5 R_{J}$ and 2.3 for all other planets, following \citet{stevenson16}. We then seek the statistical properties of the absorption, measured in scale heights.  Our study improves upon \citet{stevenson16} in several ways.  First, we increase the sample size from 14 to 34.  Also, we utilize a model atmosphere template to measure the absorption (Stevenson used absorption indices based upon restricted ranges in wavelength), and we allow for a baseline slope in the spectrum such as might be produced by small-particle scattering.  We also investigate the nature of the distribution function for exoplanetary absorption measured in scale heights. 

We use observed spectra from \citet{tsiaras17}, and derive the magnitude of the water absorption from the data as directly as possible. Essentially, we find the minimum and maximum values of the data, and convert the difference between them to scale heights of absorption.  But we must allow for the shape of the water band (absorption varies with wavelength), the scatter in the data points, and the possibility of instrumental or astrophysical baseline slopes (e.g., by small-particle scattering).  We accomplish that by calculating a nominal model spectrum for each planet, and use it as a template to gauge the amplitude of the absorption.  We calculate the equilibrium temperature of each planet, assuming zero albedo and uniform re-distribution of heat, and the surface gravity from published planetary masses and radii.  The nominal model spectrum follows from those parameters using the Exo-Transmit code \citep{kempton16}.  We used isothermal T/P profiles, with collision-induced continuous opacities as per Table~1 of \citet{kempton16}, and line opacity for water only. We then scale the nominal spectrum ($x(\lambda)$) to greater or less absorption using a multiplicative factor ($a$), and fit it to the observed spectrum ($y(\lambda)$) using an MCMC procedure ({\it emcee}, \citealp{foreman-mackey13}) with the equation $y(\lambda)=ax(\lambda)+b\lambda + c $ where $a$ is the scaling factor, $b$ is the wavelength coefficient for the baseline slope, and $c$ is a constant. 

For each planet we then take the difference between the maximum and minimum value of $R_p^2/R_s^2$ in the fitted model spectrum after removing the slope, and we divide by $2R_pH/R_s^2$, to convert the magnitude of the absorption to scale heights, $A_{H}$.  These absorption values are listed in Table \ref{fig:two_panel_fullfit}.  Although we only use the 1.3 - 1.65\,$\mu$m part of the spectrum for our statistical study (region of strongest absorption), we also tabulate the results from the full 1.1 - 1.65\,$\mu$m range in Table \ref{fig:two_panel_fullfit}. We verified that our results do not change significantly if we fit to the full 1.1 - 1.65\,$\mu$m range.

Figure \ref{fig:two_panel_fullfit} shows the fit to WASP-67b, that has the median $\chi_\nu^2$ in our sample. The posterior distributions for the $a$ and $b$ coefficients (and thus for $A_H$) are very close to Gaussian, reflecting the high quality of the HST data.  We derive the errors on $A_H$ from those posterior distributions.   

The 30 spectra presented in \citet{ tsiaras17} were derived using a uniform and consistent data analysis method \citep{tsiaras16}. However, it is still advantageous to compare them with independent spectra derived by other groups (Table \ref{table1}, right columns).  When we fit to the other spectra the same way, we derive statistically consistent absorptions in scale heights.  Although a few planets (HAT-P-1b, HD\,189733b, HD\,209458b) show some difference, we do not detect a systematic deviation. The slope (1.14$\pm$0.09) of an orthogonal distance fit \citep{akritas96} is within 2$\sigma$ of unity, indicating that spectra from \citet{tsiaras17} are consistent with those derived by other groups. We conclude that we are working with a valid collection of spectra in the sense that there are no internal inconsistencies in the measurements.\\

\begin{table*}[t]
\caption{Absorption in scale heights ($A_H$), based on spectra from \citet{tsiaras17} unless otherwise noted.  Our analysis used the $A_H$ values from fitting to the strongest region of water absorption (1.3 - 1.65\,$\mu$m), versus the entire WFC3 bandpass (1.1 - 1.65\,$\mu$m), both listed in the middle set of columns. The columns on the right give values for comparison, based on spectra from other authors. $T_{eq}$ calculated from parameters in corresponding references are listed in the left column.} 
\begin{threeparttable}
\begin{tabular}{c|c|c c|c c c}
\hline\hline 
Planet & $T_{eq}(K)$ & Absorption ($A_H$) & Absorption ($A_H$) & Absorption ($A_H$) & Absorption ($A_H$) & Reference \\
		&  & 1.1 - 1.65 $\mu m$	&	1.3 - 1.65 $\mu m$	 &  1.1 - 1.65 $\mu m$  &	1.3 - 1.65 $\mu m$\\[0.5ex] 
\hline 												
GJ 436 b	&	633	$\pm$	58	&	0.22	$\pm$	0.53	&	0.06	$\pm$	0.73	&	1.16	$\pm$	0.52	&	0.87	$\pm$	0.71	&	\citet{knutson14a}	\\
GJ 3470 b	&	692	$\pm$	101	&	0.70	$\pm$	0.30	&	0.29	$\pm$	0.41	&										\\
HAT-P-1 b	&	1320	$\pm$	103	&	1.50	$\pm$	0.33	&	1.27	$\pm$	0.35	&	2.51	$\pm$	0.35	&	2.88	$\pm$	0.42	&	\citet{wakeford13}	\\
HAT-P-3 b	&	1127	$\pm$	68	&	0.22	$\pm$	0.59	&	0.52	$\pm$	0.74	&										\\
HAT-P-11 b	&	856	$\pm$	37	&	2.96	$\pm$	0.62	&	2.31	$\pm$	0.71	&	3.21	$\pm$	0.64	&	2.70	$\pm$	0.82	&	\citet{fraine14}	\\
HAT-P-12 b	&	958	$\pm$	28	&	0.49	$\pm$	0.21	&	0.42	$\pm$	0.25	&										\\
HAT-P-17 b	&	780	$\pm$	34	&	0.47	$\pm$	0.60	&	0.27	$\pm$	0.78	&										\\
HAT-P-18 b	&	843	$\pm$	35	&	0.90	$\pm$	0.21	&	0.51	$\pm$	0.28	&										\\
HAT-P-26 b	&	980	$\pm$	56	&	2.35	$\pm$	0.26	&	1.92	$\pm$	0.31	&	2.22	$\pm$	0.18	&	1.89	$\pm$	0.20	&	\citet{wakeford17}	\\
HAT-P-32 b	&	1784	$\pm$	58	&	1.48	$\pm$	0.22	&	1.30	$\pm$	0.28	&										\\
HAT-P-38 b	&	1080	$\pm$	78	&	1.60	$\pm$	0.56	&	2.03	$\pm$	0.66	&										\\
HAT-P-41 b	&	1937	$\pm$	74	&	1.70	$\pm$	0.39	&	1.96	$\pm$	0.45	&										\\
HD149026 b	&	1627	$\pm$	83	&	0.79	$\pm$	0.48	&	1.09	$\pm$	0.56	&										\\
HD189733 b	&	1201	$\pm$	51	&	2.31	$\pm$	0.40	&	1.45	$\pm$	0.47	&	1.99	$\pm$	0.30	&	1.59	$\pm$	0.34	&	\citet{mccullough14}	\\
HD209458 b	&	1449	$\pm$	36	&	0.88	$\pm$	0.14	&	0.78	$\pm$	0.17	&	1.26	$\pm$	0.14	&	1.12	$\pm$	0.16	&	\citet{deming13}	\\
WASP-12 b	&	2580	$\pm$	146	&	1.60	$\pm$	0.23	&	1.62	$\pm$	0.31	&	2.07	$\pm$	0.36	&	2.07	$\pm$	0.36	&	\citet{kreidberg15}	\\
WASP-29 b	&	963	$\pm$	69	&	0.04	$\pm$	0.39	&	0.12	$\pm$	0.49	&										\\
WASP-31 b	&	1576	$\pm$	58	&	0.94	$\pm$	0.33	&	1.14	$\pm$	0.42	&	1.08	$\pm$	0.38	&	1.77	$\pm$	0.43	&	\citet{sing15}	\\
WASP-39 b	&	1119	$\pm$	57	&	1.27	$\pm$	0.14	&	1.22	$\pm$	0.16	&										\\
WASP-43 b	&	1374	$\pm$	147	&	1.46	$\pm$	0.43	&	0.95	$\pm$	0.46	&	1.47	$\pm$	0.45	&	0.98	$\pm$	0.51	&	\citet{kreidberg14AJ}	\\
WASP-52 b	&	1300	$\pm$	115	&	1.80	$\pm$	0.24	&	1.33	$\pm$	0.28	&										\\
WASP-63 b	&	1508	$\pm$	69	&	0.58	$\pm$	0.27	&	0.39	$\pm$	0.30	&										\\
WASP-67 b	&	1026	$\pm$	59	&	0.67	$\pm$	0.52	&	0.86	$\pm$	0.70	&										\\
WASP-69 b	&	964	$\pm$	38	&	0.62	$\pm$	0.11	&	0.65	$\pm$	0.13	&										\\
WASP-74 b	&	1915	$\pm$	116	&	0.77	$\pm$	0.40	&	0.97	$\pm$	0.45	&										\\
WASP-76 b	&	2206	$\pm$	95	&	1.35	$\pm$	0.18	&	1.62	$\pm$	0.21	&										\\
WASP-80 b	&	824	$\pm$	58	&	0.38	$\pm$	0.15	&	0.51	$\pm$	0.19	&										\\
WASP-101 b	&	1552	$\pm$	81	&	0.16	$\pm$	0.25	&	0.13	$\pm$	0.27	&										\\
WASP-121 b	&	2358	$\pm$	122	&	2.51	$\pm$	0.36	&	2.31	$\pm$	0.41	&										\\
XO-1 b	&	1196	$\pm$	60	&	2.68	$\pm$	0.66	&	3.33	$\pm$	0.76	&	2.50	$\pm$	0.56	&	3.11	$\pm$	0.72	&	\citet{deming13}	\\
WASP-17b$^{a}$	&	1632	$\pm$	126	&	0.93	$\pm$	0.33	&	0.44	$\pm$	0.35	&										\\
WASP-19b$^{b}$	&	2037	$\pm$	156	&	2.16	$\pm$	0.65	&	1.60	$\pm$	0.58	&										\\
GJ 1214b$^{c}$	&	573	$\pm$	35	&	0.11	$\pm$	0.09	&	0.05	$\pm$	0.13	&										\\
HD97658b$^{d}$	&	753	$\pm$	33	&	0.46	$\pm$	1.10	&	1.79	$\pm$	1.46	&										\\		
\hline 
\end{tabular}
\begin{tablenotes}
\item {\bf Notes.}
\small
\item ${^a}$ \citet{mandell13}
\item ${^b}$ \citet{huitson13}
\item ${^c}$ \citet{kreidberg14N}
\item ${^d}$ \citet{knutson14b}
\end{tablenotes}
\end{threeparttable}
\label{table1}
\end{table*}

\section{Statistical Correlations}

Armed with $A_H$ values from table \ref{table1}, we investigated their relationship with planetary temperature, mass, radius, and surface gravity. These relationships are subtle, and the statistics are fragile. However, we interpret them boldly, so as to form hypotheses that can stimulate and guide future work. 

\begin{figure}
\includegraphics[width=0.5\textwidth,keepaspectratio]{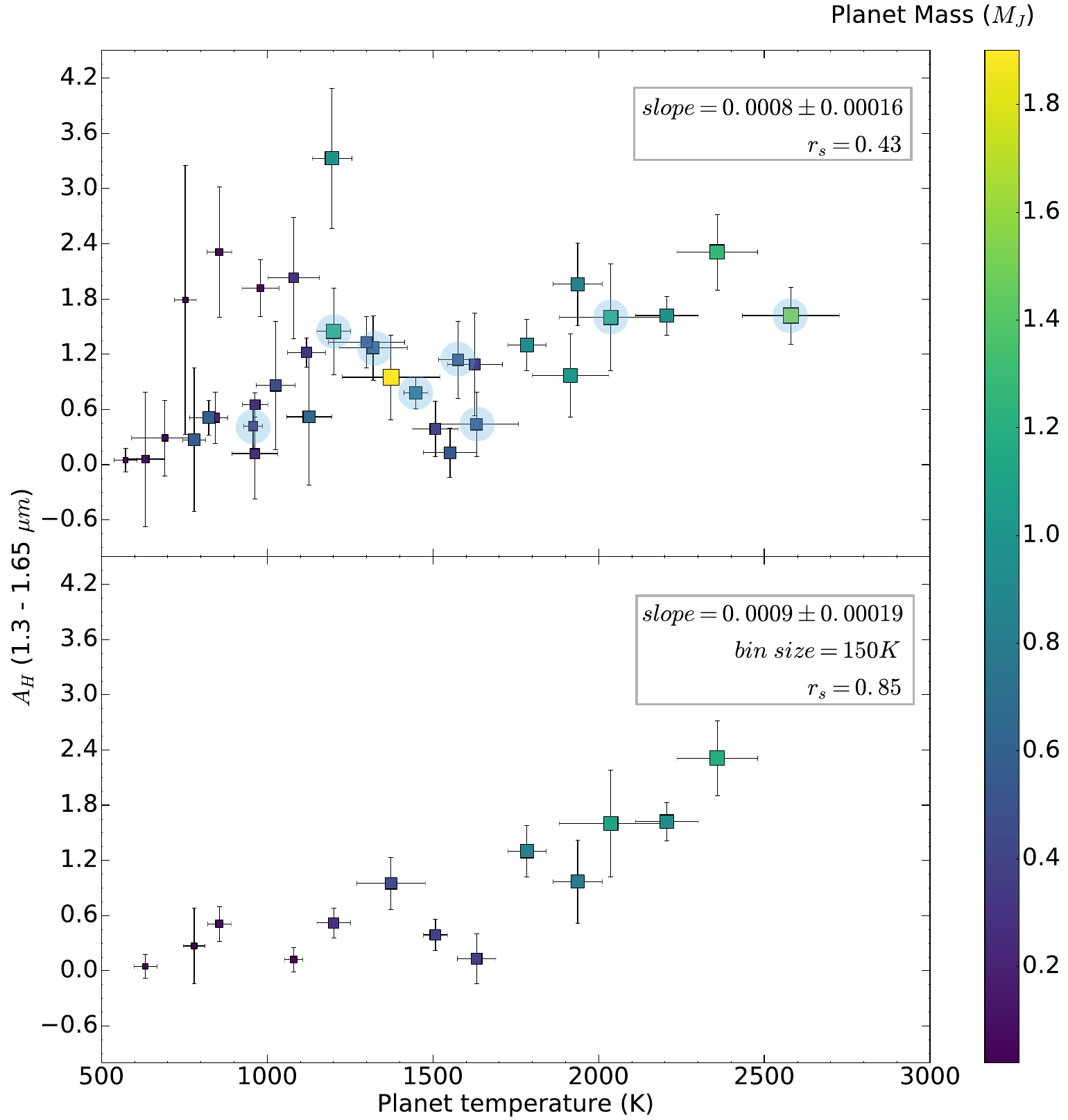}
  \caption{The upper panel is the 1.3 - 1.6\,$\mu$m absorption in scale heights versus planet equilibrium temperature. We infer a positive baseline slope correlation (p-value = 0.01) with upward scattering on the left side. After applying the binning method discussed in section 3, we obtain a statistically significant ($r_s=0.85$) baseline correlation as shown in the lower panel. Mass uncertainty is proportional to squares size. The eight planets shaded in blue circles in the upper panel are the ones investigated in \citep{sing16} }
  \label{fig:3D_temp_absorp_mass}
\end{figure}

The first point to note is that the median $A_H$ value is only 1.4, less than expected for clear solar abundance atmospheres ($A_H \approx 5$). That can be due to either clouds \citep{barstow17} or low abundance of water vapor \citep{madhu14}.  As for correlations, in the top panel of Figure \ref{fig:3D_temp_absorp_mass}, we show the relation between $A_H$ and planetary equilibrium temperature.  We propagate errors in the stellar and orbital parameters to yield errors in the abscissa as well as the ordinate. An orthogonal distance regression yields a slope of $0.0008 \pm 0.00016$, and the Spearman correlation coefficient is $0.43$, indicating a moderate correlation.  We emphasize that the temperature dependence of the atmospheric scale height has already been removed from the ordinate, so this correlation is a physical effect beyond the atmospheric scale height.  We hypothesize that the dominant effect is the decreasing amount of cloud condensation as the planetary equilibrium temperature increases. To the extent that hotter planets have fewer cloud-forming condensate species present in their atmospheres than cooler planets (\citealp{barstow17}, \citealp{kataria16}), that will tend to produce a positive slope between temperature and $A_H$.  

We explored using a mass-metallicity power law, emulating Figure~4 of \citet{kreidberg14AJ}, to calculate atmospheric molecular weight.  That has little effect on most planets in our sample because they are hot Jupiters with predominantly H-He atmospheres.  The power law causes two planets to scatter to greater $A_H$ values at the left edge of Figure~2, degrading the correlation, but not affecting the baseline derived on the lower panel of Figure~2.

A similar temperature versus $A_H$ correlation was reported by \citet{crossfield17} with a sample size comprised of six Neptune-size planets. The correlation led them to suggest that hazes might become more significant for planets with $T_{eq} < 850 K$. Our study includes their six planets and also an additional six planets with  $T_{eq} < 1000 K$. We do not see a clear divide of $A_H$ values around $T_{eq} = 850 K$ as shown by \citet{crossfield17}. However, our six additional planets are not Neptune-like but rather more massive ($\sim$0.2 to $\sim$0.5 $M_J$) planets. Another in-depth study conducted by \citet{sing16} used eight planets with extensive wavelength coverage from HST/Spitzer transmission spectra.  Although those eight planets have $T_{eq}$ranging from $\sim$1000K to $\sim$2500K, \citet{sing16} found no trend between and $T_{eq}$ and the magnitude of water absorption. In the upper panel of \ref{fig:3D_temp_absorp_mass}, we shaded the eight planets from \citep{sing16} with blue circles. Those eight planets are not sufficient to establish a clear correlation (p-value of 0.3 for a linear trend).

Another effect that may be present on Figure \ref{fig:3D_temp_absorp_mass} is a "baseline" value for $A_H$ at each equilibrium temperature, with scatter above that baseline value, especially for equilibrium temperatures below 1500K. To better characterize this baseline effect, we developed a binning analysis method that divides the data according to a chosen bin size and takes the lowest point in each bin. This way, upward scattering points will be filtered out and only the points that form the baseline will remain. However, the resulting baseline correlation from this method will depend on the chosen bin size. To support the validity of this binning method and find the optimal bin size, we tested it on randomly generated absorption values. We averaged the results from 1000 runs of random values and compared with real absorption data.  This test indicated that false baseline correlations can result from binning random data, but those correlations are much weaker than we find when we bin the real data. Using a 150K bin size on the real data, we obtained a baseline slope with very strong positive correlation ($R_s$=0.85) as shown in lower plot in Figure \ref{fig:3D_temp_absorp_mass}. Binning random data with this bin size produces only $R_s$=0.35, a weak effect.  Using orthogonal distance regression on the binned real data, we find a slope of 0.0009 $K^{-1}$. Thus, for each 1000K increase in planetary equilibrium temperature, we find that the baseline (i.e., statistical minimum) water vapor absorption increases by about 0.9 scale heights.

The top panel of Figure \ref{fig:3D_temp_absorp_mass} shows that the cooler planets tend to scatter above our inferred baseline.  This could be due to variable cloud coverage, variable water vapor abundance, or variations in surface gravity that cause different cloud distributions at a given equilibrium temperature. Surface gravity is the most amenable to investigation, and we now turn to that possible correlation. 

\begin{figure*}
  \includegraphics[width=1\textwidth]{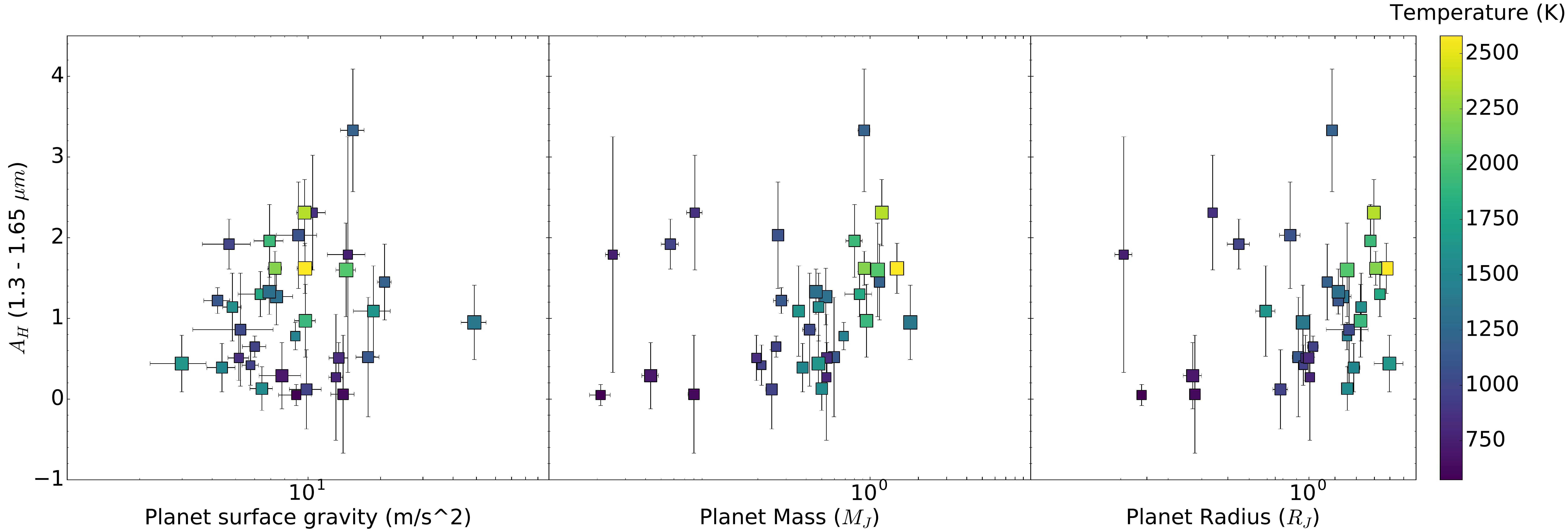}
  \caption{Planet surface gravity, mass, and radius versus absorption in scale heights (1.3-1.65\,$\mu$m). We do not detect any significant statistical correlation; the suggestion of a correlation with mass is due to an intrinsic degeneracy in the sample, combined with the temperature correlation shown in Figure~2.}
  \label{fig:figure_3}
\end{figure*}

Figure \ref{fig:figure_3} plots $A_H$ versus surface gravity, planet mass and radius. We do not see any clear statistical correlation between surface gravity, or radius, and $A_H$.  There is an apparent correlation with mass, but it is due to the temperature correlation  (Figure~2), combined with an intrinsic degeneracy in the sample, as discussed below.  The fact that increasing surface gravity does not seem to have a significant effect on $A_H$ reinforces the expectation that surface gravity is not directly linked to the cloud formation process. In simple models of cloud formation \citep{burrows99}, it is largely a chemical process determined by the equilibrium temperature and thermal structure of the planet. Our statistical results are consistent with that paradigm.

\begin{figure}
  \includegraphics[width=0.5\textwidth,keepaspectratio]{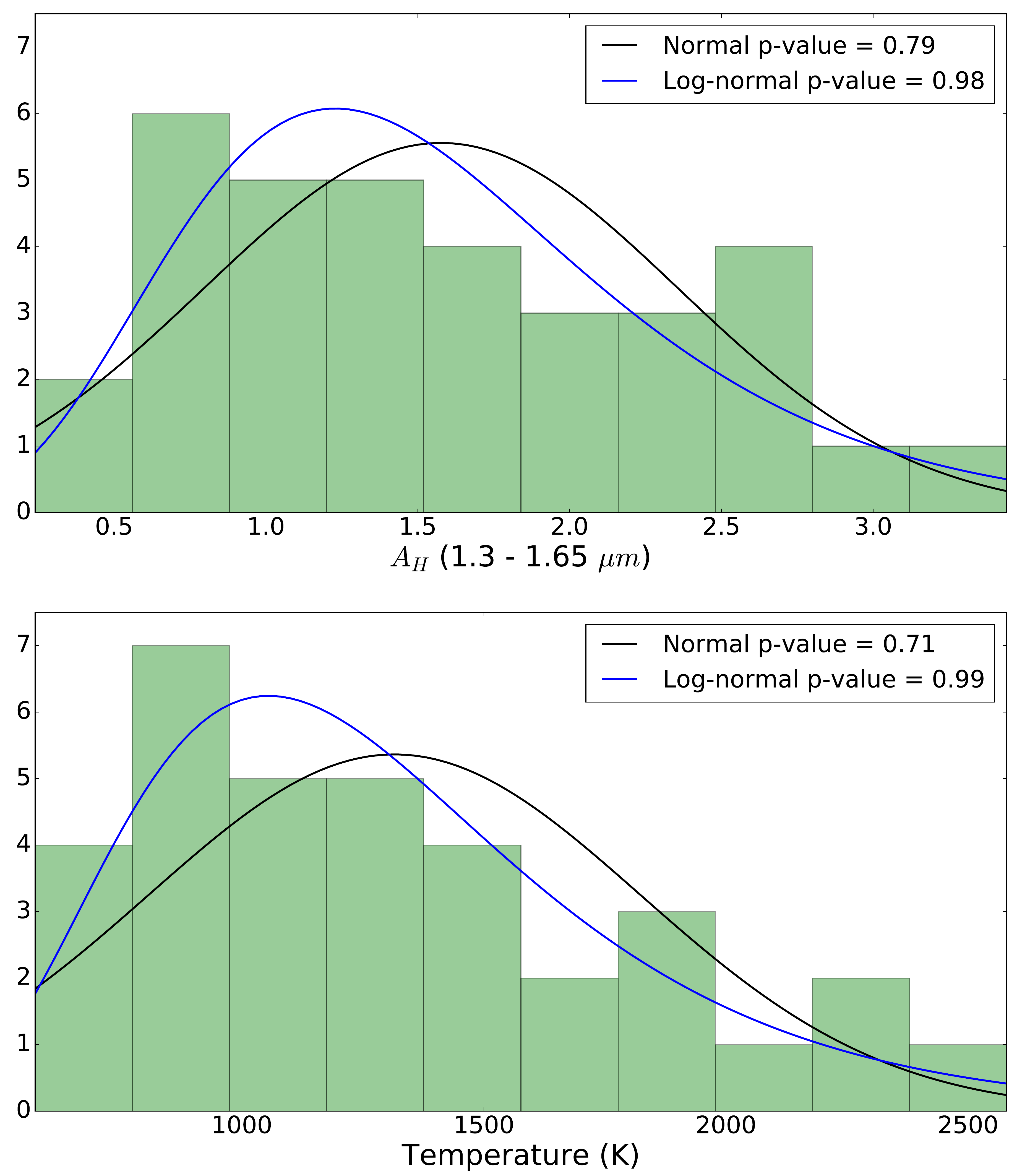}
  \caption{Upper plot shows the distribution of absorption in scale heights (1.3 - 1.65\,$\mu$m) is more likely to be log-normal than normal. This is caused by the target selection bias as there are more cooler planets than hotter ones in the sample (shown in the lower panel). }
  \label{fig:Distribution}
\end{figure}

We also investigated the distribution function of $A_H$ values. We noticed that the most likely distribution of $A_H$ is not Gaussian but rather log-normal as shown in the upper panel in Figure \ref{fig:Distribution}. In principle, this could reflect the complexity of transit absorption, since log-normal distributions usually result when the underlying processes are multiplicative as opposed to additive.  However, we think this is likely due to a target selection bias instead of fundamental physical processes, since the  distribution of equilibrium temperatures (lower panel in Figure \ref{fig:Distribution}) similarly favors a log-normal distribution. Evidently, observers have been favoring cool planets as opposed to hot ones.  We suggest that more hot planets should be included in future observations to ensure unbiased samples (i.e., more closely approaching a uniform distribution).\\

\section{Summary and Conclusions}

We have expanded the sample studied by \citet{stevenson16} from 14 to 34 planets ranging from super-Earths to hot Jupiters.  We used Exo-Transmit templates to fit the observed spectra, and express the water absorption in units of scale heights ($A_H$), removing known physical dependencies. Comparing with results from \citet{stevenson16}, we find one continuous positive correlation between $A_H$ and $T_{eq}$ ranging from $\sim$500K to $\sim$2500K as opposed to a strong correlation only when $T_{eq}$ < 750K. \citet{stevenson16} also reported a weak correlation between surface gravity and $A_H$, but we see no clear correlation between those two parameters.  Our results are qualitatively consistent with the temperature correlation inferred by \citet{crossfield17} for Neptune-like planets. 	

We point out that the observed sample of exoplanets (Table \ref{table1}) contains an intrinsic degeneracy in the sense that planetary mass is correlated with equilibrium temperature (correlation coefficient = 0.75, a strong correlation).  Moreover, our division by the scale height is equivalent to multiplying by mass.  In principle, it is possible for the temperature correlation to be created as an artifact of our analysis method. However, that would require that the total observed water absorption is not proportional to the atmospheric scale height, a very unlikely condition.  Therefore we consider the correlation between $T_{eq}$ and $A_H$ as real and physical.   

The $A_H$ vs. $T_{eq}$ correlation could be caused by physical mechanisms such as cloud formation and longitudinal circulation of heat. \citet{barstow17} discussed cloud formation as a continuum process based on atmospheric thermal structure. At cooler atmospheric temperatures, clouds fall deeper while new species condense in the upper atmosphere. This process will naturally leave cooler planets with more extended cloud coverage than hotter planets.  Also, planetary heat circulation has been shown to be inefficient for hotter planets \citep{fortney08, cowan11}.  This means that for the hottest planets, the terminator regions we probe through transmission spectroscopy are likely to be cooler than our equilibrium temperature, and the sub-stellar regions hotter than our equilibrium temperature. Consequently, our calculated scale heights for the hottest planets are arguably too large, and the true scale heights would further strengthen the $A_H$-temperature correlation and increase the baseline slope.

Unfortunately, with the quality and sample size of current spectra, degeneracies between clouds, temperature and mean molecular weight can not yet be resolved. However, we favor the cloud interpretation because we deem it to be the most physically based and plausible explanation.  We infer a selection bias that prefers cooler planets in all the targets observed to date.  A greater proportion of the hottest planets, especially at low mass, should be included in future observations to better constrain the correlations, and also to break the mass-temperature degeneracy in the current sample.
\vfill
We thank an anonymous referee and also Dr. Nikole Lewis and Dr. Hannah Wakeford for insightful comments that helped us improve this paper.


\begin{thebibliography}{}
\bibitem[Akritas \& Bershady (1996)]{akritas96} Akritas, M. G., \& Bershady, M. A. 1996, ApJ, 470, 706
\bibitem[Barstow et al. (2017)]{barstow17} Barstow, J. K., Aigrain, S., Irwin, P. G. J., \& Sing, D. K. 2017, ApJ, 834, 50
\bibitem[Burrows \& Sharp (1999)]{burrows99} Burrows, A., \& Sharp, C. M. 1999, ApJ, 512, 843
\bibitem[Cowan \& Agol (2011)] {cowan11} Cowan, N.~B., \& Agol,~E. 2011, ApJ, 729, id.54.
\bibitem[Crossfield(2015)] {crossfield15} Crossfield,~I.~J.~M. 2015, PASP, 127, 941.
\bibitem[Crossfield \& Kreidberg (2017)]{crossfield17} Crossfield, I., Kreidberg, L. 2017, ArXiv e-prints, arXiv:1708.00016
\bibitem[Deming et al. (2013)]{deming13} Deming, D., Wilkins, A., McCullough, P., et al. 2013, ApJ, 774, 95
\bibitem[Deming \& Seager (2017)]{deming17} Deming D., Seager S., 2017, JGRE , 122, 53
\bibitem[Evans et al.(2017)] {evans17} Evans, T.~M., et al. 2017, Nature, 548, 58.
\bibitem[Foreman-Mackey et al.(2013)]{foreman-mackey13} Foreman-Mackey,~D., Hogg,~D.~W., Lang,~D., \& Goodman,~J. 2013, PASP, 125, 306.  
\bibitem[Fortney et al. (2008)]{fortney08}Fortney, J. J., Lodders, K., Marley, M. S., \& Freedman, R. S. 2008, ApJ, 678, 1419
\bibitem[Fraine et al. (2014)]{fraine14} Fraine, J., Deming, D., Benneke, B., et al. 2014, Nature, 513, 526
\bibitem[Griffith (2014)]{griffith14} Griffith, C. A. 2014, Philosophical Transactions of the Royal Society of London Series A, 372, 30086
\bibitem[Haynes et al.(2015)]{haynes15} Haynes, K., Mandell, A. M., Madhusudhan, N., Deming, D., Knutson, H. 2015, ApJ, 806, 146
\bibitem[Heng et al. (2017)]{heng17} Heng, K., Kitzmann, D., 2017, MNRAS, 2017, stx1453. doi: 10.1093/mnras/stx1453
\bibitem[Huitson et al. (2013)]{huitson13} Huitson, C. M., Sing, D. K., Pont, F., et al. 2013, MNRAS, 434, 3252
\bibitem[Kataria et al. (2016)]{kataria16} Kataria, T., Sing, D. K., Lewis, N. K., et al. 2016, ApJ, 821, 9
\bibitem[Kempton et al. (2016)]{kempton16} Kempton, E. M.-R., Lupu, R. E., Owusu-Asare, A., Slough, P., \& Cale, B. 2016, arXiv:1611.03871
\bibitem[Knutson et al. (2014a)]{knutson14a} Knutson, H. A., Benneke, B., Deming, D., \& Homeier, D. 2014a, Nature, 505, 66
\bibitem[Knutson et al. (2014b)]{knutson14b} Knutson, H. A., Dragomir, D., Kreidberg, L., et al. 2014, ApJ, 794, 155
\bibitem[Kreidberg et al. (2014a)]{kreidberg14AJ} Kreidberg, L., Bean, J. L., Désert, J.-M., et al. 2014, ApJL, 793, L27
\bibitem[Kreidberg et al. (2014b)]{kreidberg14N} Kreidberg, L., Bean, J. L., Désert, J.-M., et al. 2014, Natur, 505, 69
\bibitem[Kreidberg et al. (2015)]{kreidberg15} Kreidberg, L., Line, M. R., Bean, J. L., et al. 2015, ApJ, 814, 66
\bibitem[Line et al. (2016)]{line16a} Line, M. R., Stevenson, K. B., Bean, J., D´esert, J.-M., Fortney, J. J., Kreidberg, L., Madhusudhan, N., Showman, A. P., \& Diamond-Lowe, H. 2016, AJ, 152, 203
\bibitem[Line et al. (2016)]{line16b} Line, M.R., Parmentier, V., 2016, ApJ, 820, 78
\bibitem[MacDonald \& Madhusudhan (2017)]{macdonald17} MacDonald R. J., Madhusudhan N., 2017, MNRAS469, 1979
\bibitem[Madhusudhan et al. (2014)]{madhu14} Madhusudhan, N., Crouzet, N., McCullough, P. R., Deming, D., \& Hedges, C. 2014a, ApJL, 791, L9
\bibitem[McCullough et al. (2014)]{mccullough14} McCullough, P. R., Crouzet, N., Deming, D., \& Madhusudhan, N. 2014, ApJ, 791, 55 
\bibitem[Mandell et al. (2013)]{mandell13} Mandell, A. M., Haynes, K., Sinukoff, E., et al. 2013, ApJ, 779, 128
\bibitem[Sing et al. (2016)]{sing16} Sing, D. K., Fortney, J. J., Nikolov, N., et al. 2016, Natur, 529, 59	
\bibitem[Sing et al. (2015)]{sing15} Sing, D. K., Wakeford, H. R., Showman, A. P., et al. 2015, MNRAS, 446, 2428
\bibitem[Stevenson et al. (2017)]{stevenson17} Stevenson, K. B., Line, M. R., Bean, J. L., et al. 2017, AJ, 153, 68
\bibitem[Stevenson (2016)]{stevenson16} Stevenson, K. B. 2016, ApJL, 817, L16
\bibitem[Stevenson et al. (2014)]{stevenson14} Stevenson, K. B., Desert, J.-M., Line, M. R., et al. 2014, Sci, 346, 838	
\bibitem[Thorngren et al. (2016)]{thorngren16} Thorngren, D. P., Fortney, J. J., Murray-Clay, R. A., \& Lopez, E. D. 2016, ApJ, 831, 64 
\bibitem[Tsiaras et al. (2017)]{tsiaras17} Tsiaras, A., Waldmann, I. P., Zingales, T., et al. 2017, ArXiv e-prints, arXiv:1704.05413	
\bibitem[Tsiaras et al. (2016)]{tsiaras16} Tsiaras, A., Rocchetto, M., Waldmann, I., et al. 2016, ApJ, 820, 99
\bibitem[Wakeford et al. (2013)]{wakeford13} Wakeford, H. R., Sing, D. K., Deming, D., et al. 2013, MNRAS, 435, 3481
\bibitem[Wakeford et al. (2017)]{wakeford17} Wakeford, H. R., Sing, D. S., Kataria, T., et al. 2017 Science, 356, 628	

\end{thebibliography}
\end{document}